\documentclass[10pt,a4paper,twoside]{article}
\usepackage{epsfig}
\usepackage{baltlat6}
\usepackage{array}
\usepackage{here}
\usepackage[
singlelinecheck=false 
]{caption}
\begin{document}
\ \
\vspace{0.5mm}
\setcounter{page}{221}

\titlehead{Baltic Astronomy, vol.\,23, 221--229, 2014}

\titleb{MODELS OF LATE-TYPE DISK GALAXIES: 1-D VERSUS 2-D}

\begin{authorl}
\authorb{T. Mineikis}{1,2} and
\authorb{V. Vansevi\v{c}ius}{1,2}
\end{authorl}

\begin{addressl}
\addressb{1}{Vilnius University Observatory, \v{C}iurlionio 29,
Vilnius LT-03100, Lithuania
vladas.vansevicius@ff.vu.lt}
\addressb{2}{Center for Physical Sciences and Technology,
 Savanori\c{u} 231, Vilnius LT-02300, Lithuania
tadas.mineikis@ftmc.lt}
\end{addressl}

\submitb{Received: 2014 November 20; accepted: 2014 December 23}

\begin{summary} 
We investigate the effects of stochasticity on the observed galaxy
parameters by comparing our stochastic star formation two-dimensional (2-D)
galaxy evolution models with the commonly used one-dimensional (1-D) models
with smooth star formation. The 2-D stochastic models predict high variability
of the star formation rate and the surface photometric parameters across the
galactic disks and in time.
\end{summary}

\begin{keywords} galaxies: general -- galaxies: evolution \end{keywords}

\resthead{Models of late-type disk galaxies: 1-D versus 2-D}
{T. Mineikis, V. Vansevi\v{c}ius}

\sectionb{1}{INTRODUCTION}

The studies of well resolved nearby disk galaxies demonstrate bursting nature of
star formation (SF) events stochastically distributed across the galaxy disks (e.g.,
Harris \& Zaritsky 2009). The effects of bursting SF are the most prominent in
the outer disk regions (e.g., Barnes et al. 2013), however, photometric parameters,
especially in the UV spectral region, are significantly altered throughout the entire
disks (e.g., Simones et al. 2014). Despite that, the one-dimensional (1-D, resolved
only along the radius of a galaxy) models, used for simulation of radial variations
of the parameters of a disk galaxy, still employ a smooth SF prescription (e.g.,
Magrini et al. 2007; Marcon-Uchida et al. 2010; Kang et al. 2012).

In our previous paper (Mineikis \& Vansevi\v{c}ius 2014), we used two-dimensional
(2-D, resolved along the radius and azimuth of a galaxy) models downgraded to
1-D profiles and compared with the observed radial profiles of the galaxy M33.
In this case, the presence of the so-called parameter ``degeneracy valley" of 2-D
models was demonstrated. In order to break those degeneracies, the necessity to
use additional 2-D observational information was stressed.

In this study we explore the 2-D galaxy models with stochastic SF, which re-
produce satisfactorily the observed radial profiles of the late-type spiral galaxy
M33 (Mineikis \& Vansevi\v{c}ius 2010, 2014). We use the 2-D models, with the parameters adjusted for M33, and compare them with the corresponding 1-D models
with smooth SF in order to reveal SF stochasticity effects.

%

\begin{table}[H]
\vbox{\tabcolsep=4pt
\parbox[c]{124mm}{\baselineskip=0pt
{\smallbf\ \ Table 1.}{ PEGASE-HR parameters}}
\center
\begin{tabular}{l l l}      
\hline              
Parameter &  Value  & Reference \\ 
\hline                       
Stellar library              & low-resolution & Le Borgne et al. (2004) \\
Initial mass function        & corrected for binaries  & Kroupa (2002) \\
Fraction of close binaries	 & 0.05 & default PEGASE-HR value \\
Ejecta of massive stars	     & type B & Woosley \& Weaver (1995) \\
Nebular emission             & true & PEGASE-HR value  \\
\hline
\end{tabular}
}
\end{table}
\vskip-5mm
\sectionb{2}{THE MODELS}

\subsectionb{2.1}{Smooth SF 1-D model}

We construct 1-D galaxy disk model following the approaches developed for
the nearby disk galaxies (e.g., Marcon-Uchida et al. 2010). The model is defined
by dividing disk into concentric rings of 1 kpc in width. The disk grows by a
gradual gas accretion from the reservoirs (attributed to each ring), where gas
resides initially. The gas does not migrate in radial direction and always is well
mixed. The individual evolution of each ring is computed by using the package
PEGASE-HR (Le Borgne et al. 2004) and applying the Schmidt-Kennicutt type
SF law (Kennicutt 1998):

\begin{equation}
{\rm SFR}_{i}=\frac{1}{\tau_{{\rm SF},i}}
\cdot\left( \frac{\Sigma_{{\rm G},i}(t)}{\Sigma_{0,i}} \right)^{n},
\end{equation}
where SF rate in the ring $i$, ${\rm SFR}_{i}$, is proportional to the gas density $\Sigma_{ {\rm G},i}$ normalized by the initial reservoir mass density $\Sigma_{0,i}$; the star formation parameter $\tau_{{\rm SF},i}$ defines the time-scale of the galaxy ring build up. The PEGASE-HR parameters
used to generate 1-D and 2-D models are given in Table 1.

\subsectionb{2.2}{Stochastic SF 2-D model}

We construct 2-D galaxy disk model following the prescription given in Mineikis \& Vansevi\v{c}ius (2014). The disk is building up by a gradual gas accretion from
the reservoir, where gas resides initially. The SF in the model is simulated by
stochastic SF events in the model cells. The main model parameters determining
stochastic SF are the probability of triggered SF, $P_{\rm T}$, and the efficiency of SF, $\epsilon$ and $\alpha$. $P_{\rm T}$ controls the intensity of propagating SF. The SF efficiency depends on
two parameters and a gas density in a cell:

\begin{equation}
{\rm SFE} = \epsilon\cdot \left( \frac{\Sigma_{\rm G}}{10\,
 M_{\odot}/{\rm pc^2}} \right)^{\alpha}
\end{equation}

where $\Sigma_{\rm G}$ is the gas surface density in a cell.

\subsectionb{2.3} {Calibration of the models}

For the comparison with the 1-D models we use two different 2-D models taken
from the parameter ``degeneracy valley" (Mineikis \& Vansevi\v{c}ius 2014). The 2-D
models along the ``degeneracy valley" have, on average, the radial profiles of gas
surface density and \textit{i} band photometry indistinguishable within 10\% of accuracy.
However, SF stochasticity (Fig. 1) and clustering of the SF regions (see Fig. 5
in Mineikis \& Vansevi\v{c}ius 2014) are significantly different among them. The SF parameters of 2-D models used for the comparison with 1-D models are given in Table 2.

\begin{figure}[tH]
\vbox{
\centerline{\psfig{figure=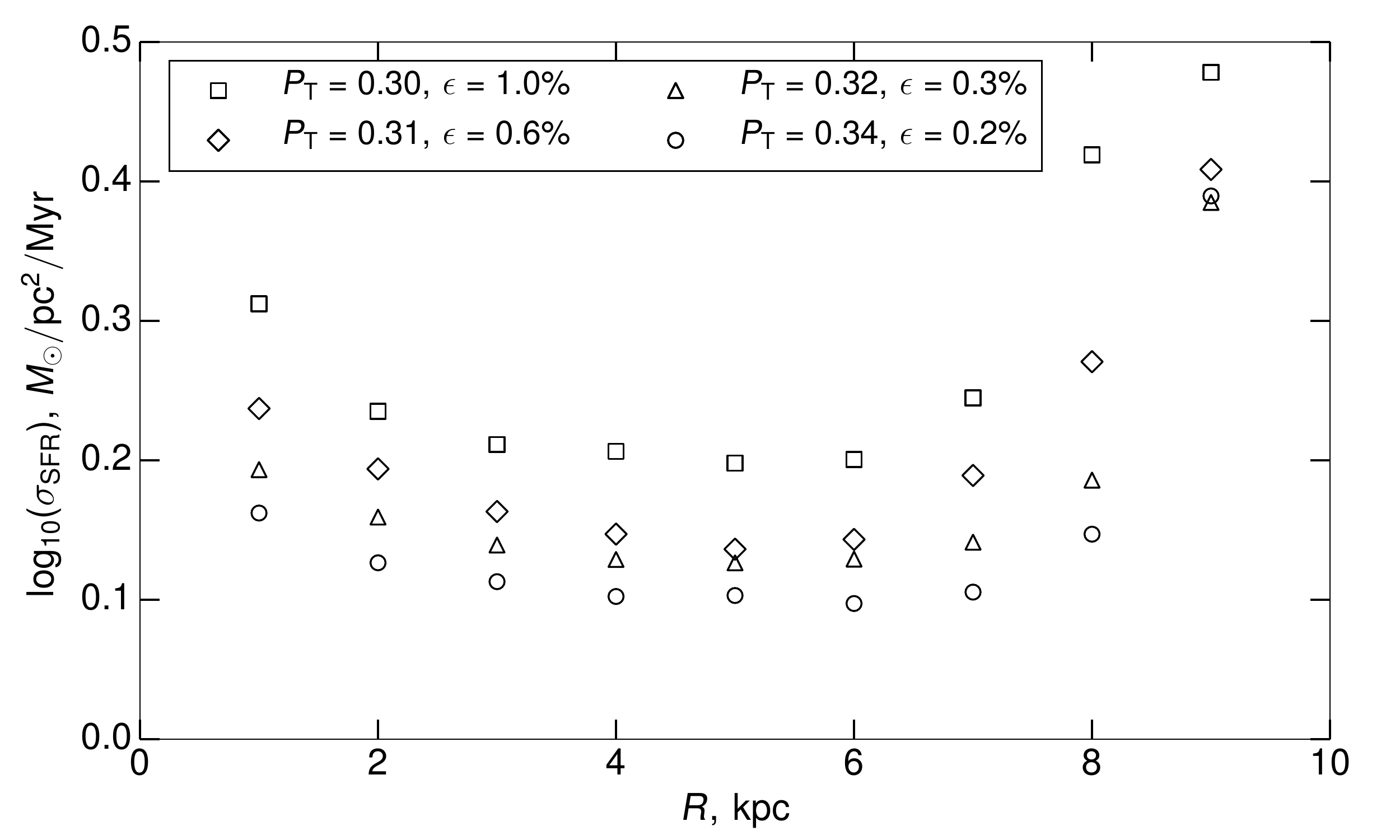,width=100mm,angle=0,clip=}}
\vspace{1mm}
\captionb{1}
{Radial profiles of the SFR standard deviation of the 2-D models.
The SFR standard deviation was calculated in radial bins of 1 kpc in
width for the galaxy ages from 10 to 13\,Gyr
with a time step of 10\,Myr.}
}
\end{figure}
\begin{center}
%
\begin{table}[tH]
\vbox{\tabcolsep=8pt
\parbox[c]{124mm}{\baselineskip=10pt
\center
{\smallbf\ \ Table 2.}{ The parameters of 2-D models}}
\center
\vskip0mm
\begin{tabular}{l l l l}       
\hline                
Model &  $P_{\rm T}$ & $\epsilon$ & $\alpha$ \\  
\hline                        
High stochasticity (HS)	     & 0.30 &    1.0\%  & 2 \\
Low stochasticity (LS)	 & 0.34 &    0.2\% & 2 \\
\hline                                   
\end{tabular}
}
%
\vbox{\tabcolsep=8pt
\parbox[c]{124mm}{\baselineskip=10pt
\center
{\smallbf\ \ Table 3.}{ The parameters of 1-D models}}
\center
\vskip-1mm
\begin{tabular}{l l l}       
\hline              
\textit{R}, kpc & ${\rm log}_{10}(\tau_{\rm SF})$, Myr & $n$\\ 
\hline                        
1.5  & 1.6 &  3 \\
2.5  & 2.1 &  3 \\
4.5  & 3.0 &  3 \\
8.5  & 4.9 &  3 \\
\hline                                   
\end{tabular}
}
\end{table}
\end{center}
\vskip-12mm

To compare 1-D and 2-D model predictions, we derived SF prescription for the
1-D models from the 2-D models fitted for the M33 galaxy. The 1-D model SF
is based on the Schmidt-Kennicutt law and is controlled via two parameters, $n$
and $\tau_{\rm SF}$. We used the parameter $n = 3$ derived from the observations (Heyer et
al. 2004). The $\tau_{{\rm SF},i}$ parameter values for each ring were derived from the 2-D LS
models by equation (1).

The evolution of $\tau_{{\rm SF},i}$ in the selected rings of the 2-D model is shown in Fig. 2.
At the start of the galaxy simulation, the values of $\tau_{{\rm SF},i}$ tend to increase in all
rings. This is due to slow SF rate because of low gas density in the cells, $\Sigma_{\rm G} < \Sigma_{\rm C}$
(for SF, the gas surface density threshold $\Sigma_{\rm C}=8\,M_{\odot}/\rm{pc^2}$). The gas density grows
faster in the inner parts of the disk, therefore, the normal SF there occurs on a
shorter time-scale. The derived median values of $\tau_{{\rm SF},i}$ for each ring are given in
Table 3.

\begin{figure}[!tH]
\vskip-5mm
\vbox{
\centerline{\psfig{figure=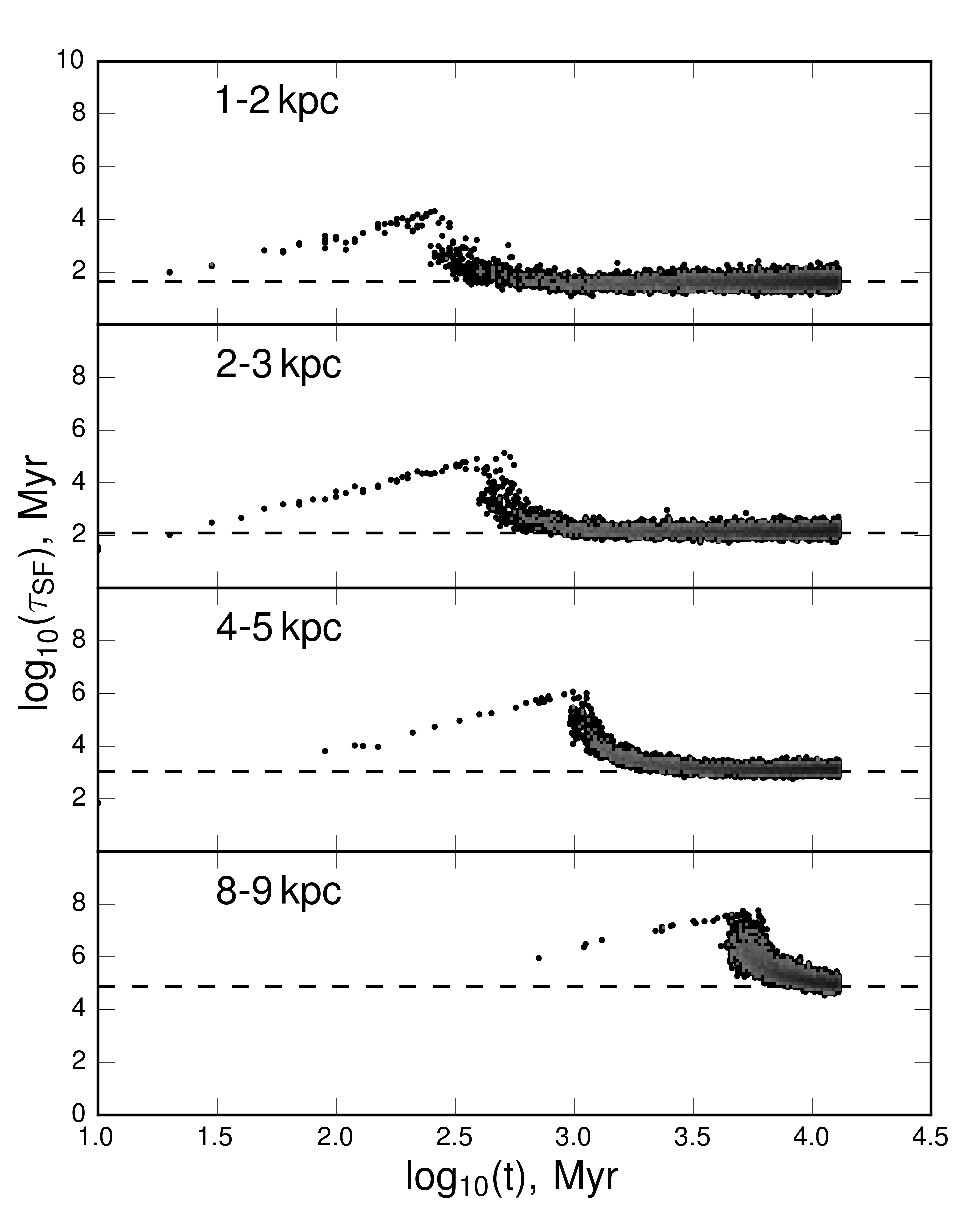,width=96mm,angle=0,clip=}}
\vspace{1mm}
\captionb{2}
{SF time-scale parameter $\tau_{{\rm SF},i}$ derived from eight 2-D models with a time step
of 10 Myr in the 1\,kpc wide rings. The dashed line indicates median values of $\tau_{{\rm SF},i}$
derived for the settled SF rate.}
}
\end{figure}

\sectionb{3}{RESULTS}

We succeeded to calibrate the smooth SF 1-D models and transform them to
the system common to the stochastic SF 2-D models. Therefore, by comparing
the evolution of the main SF parameters for both models, we were able to clearly
demonstrate the large effects arising due to SF stochasticity, see Fig.~3 for the case
of low stochasticity (LS) model.

We do not see strong stochasticity effects on the gas surface density and metallicity evolution. However, the effect of these parameters on the SF rate at the early
ages is evidenced by strong discrepancy between the 1-D and 2-D models. This
effect appears due to a low gas surface density, which is insufficient for normal
self-propagating SF in the 2-D models. A huge scatter in the SF rate is prominent, being the largest in the central regions and outskirts of the galaxy disk. In
the central disk regions the scatter is partly caused by a smaller number of cells
forming the ring 1~kpc wide. In the disk outskirts the scatter is large because the
low gas density prevents the steady self-propagating SF.

\begin{figure}[H]
\vbox{
\centerline{\psfig{figure=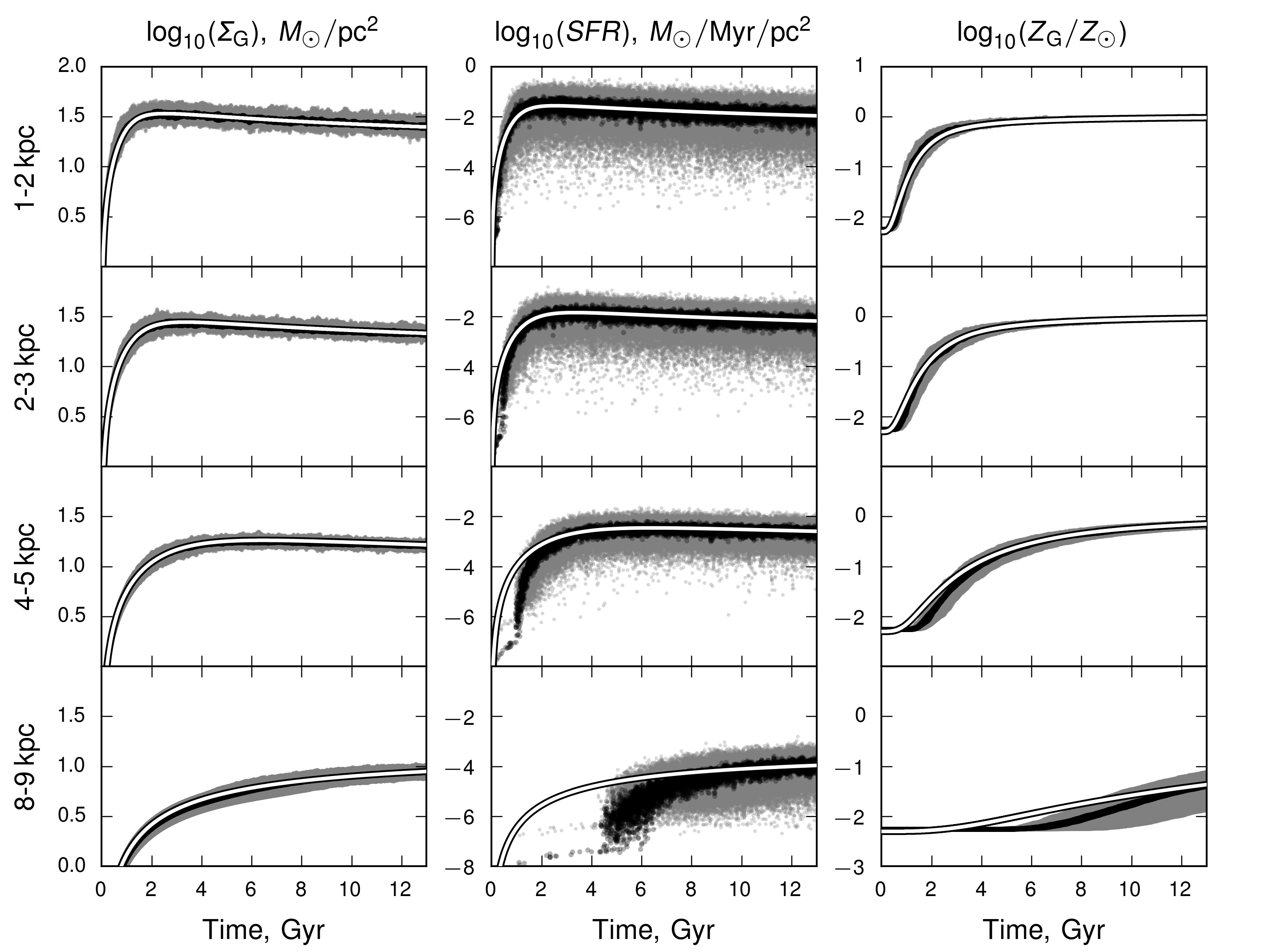,width=120mm,clip=}}
\vspace{1mm}
\captionb{3}
{The evolution of the gas surface density, SF surface density and gas metallicity
in radial rings. The white line indicates a smooth SF in 1-D galaxy model based on the
stochastic 2-D model (LS). The black dots represent eight independent runs of the LS
model averaged within 1\,kpc wide rings, the gray dots are for 0.1\,kpc wide rings. The
time step of the LS models is 10 Myr.}
}

\vbox{
\centerline{\psfig{figure=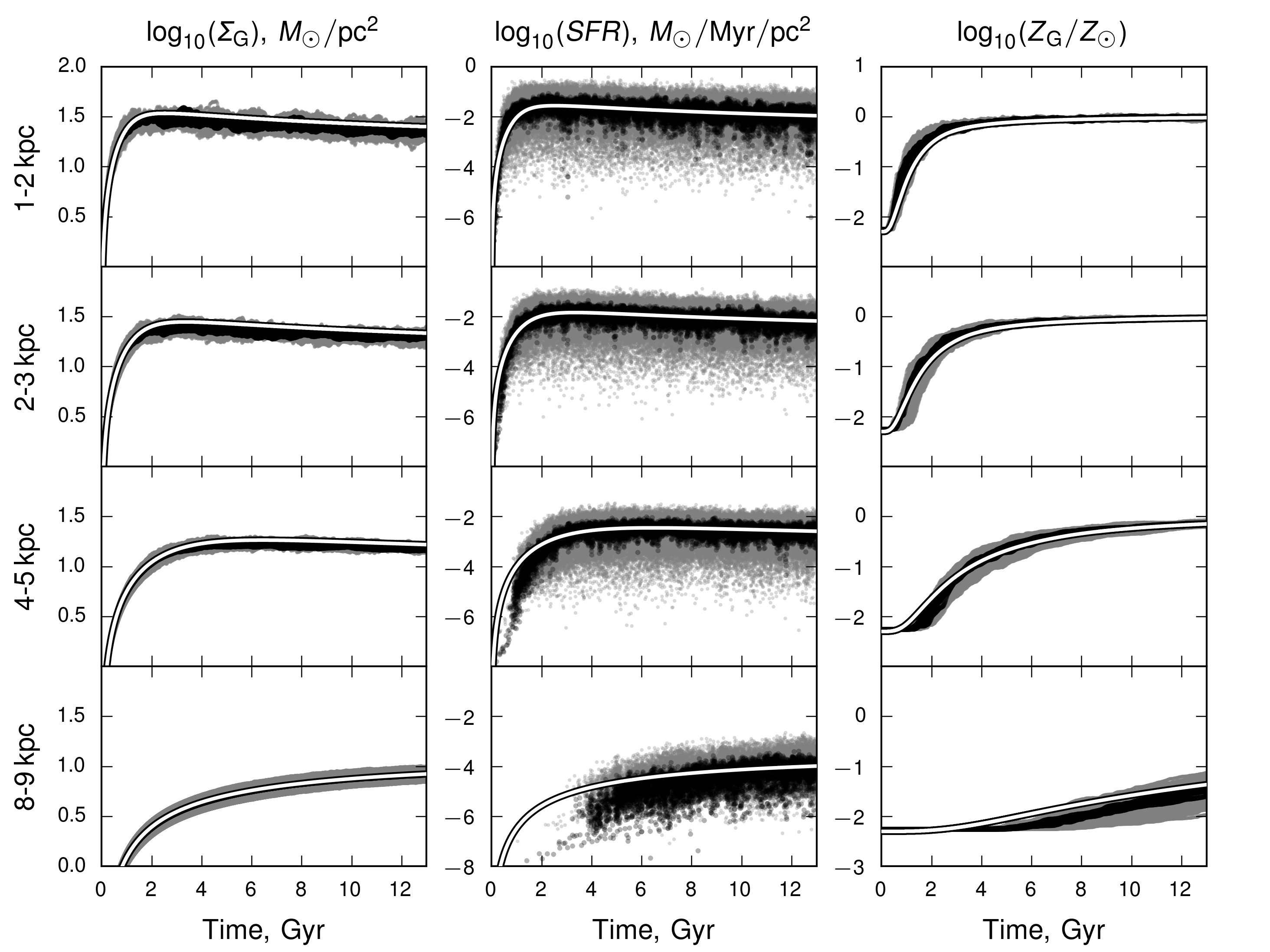,width=120mm,angle=0,clip=}}
\vspace{1mm}
\captionb{4}
{The same as in Fig.\,3, but for the HS model.}
}
\end{figure}


\begin{figure}[H]
\vbox{
\centerline{\psfig{figure=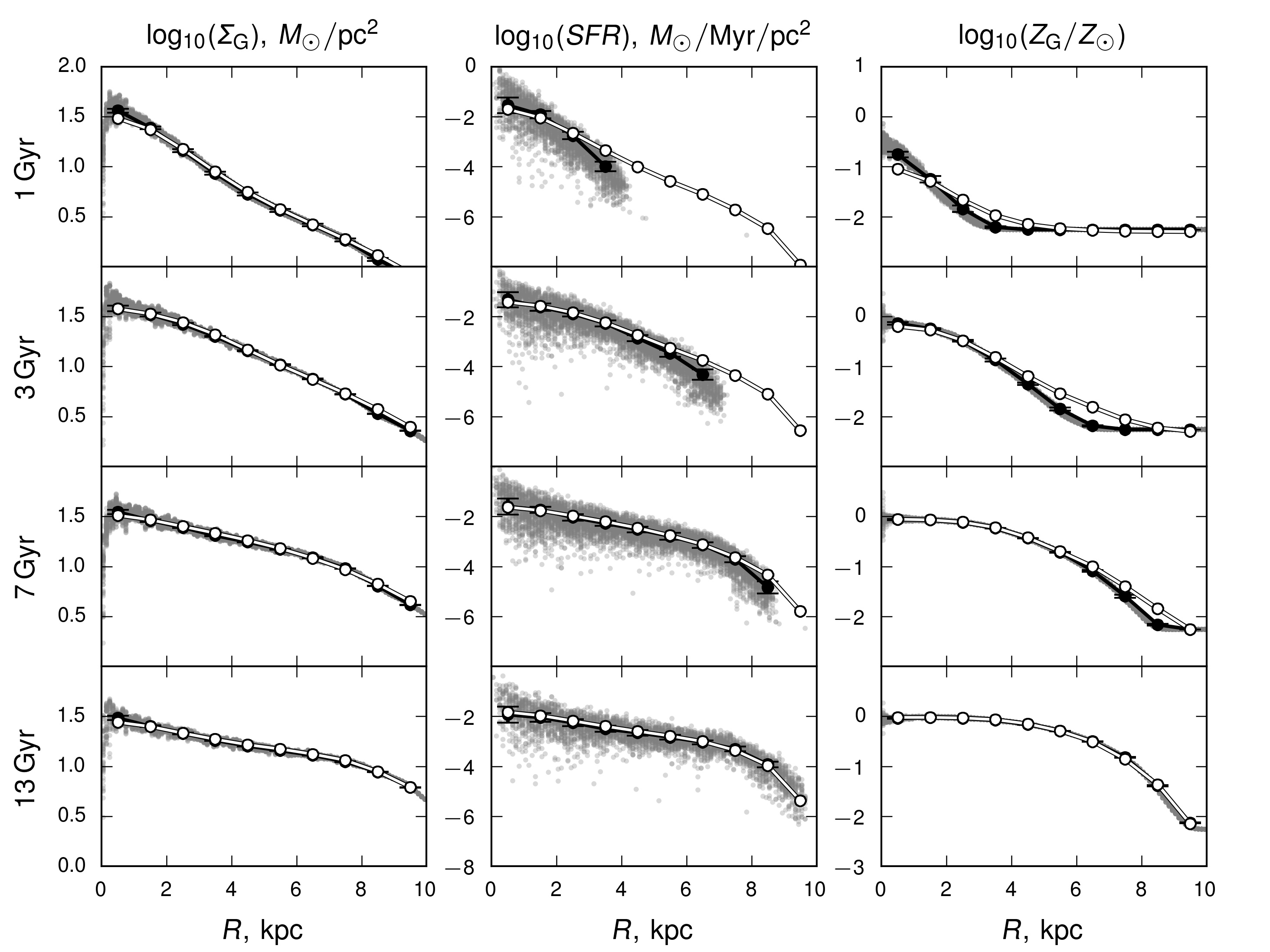,width=120mm,angle=0,clip=}}
\vspace{1mm}
\captionb{5}
{Radial profiles of the gas surface density, SF surface density and metallicity
at different time. The white circles indicate a smooth SF 1-D model. The black dots
indicate the mean values of eight independent runs of the LS model averaged within
1\,kpc wide rings; the error bars show the standard deviations in the scatter of the model
values. The gray dots indicate the LS models averaged within 0.1\,kpc wide rings. The
time bin for averaging was 100\,Myr.}
}
\vbox{
\centerline{\psfig{figure=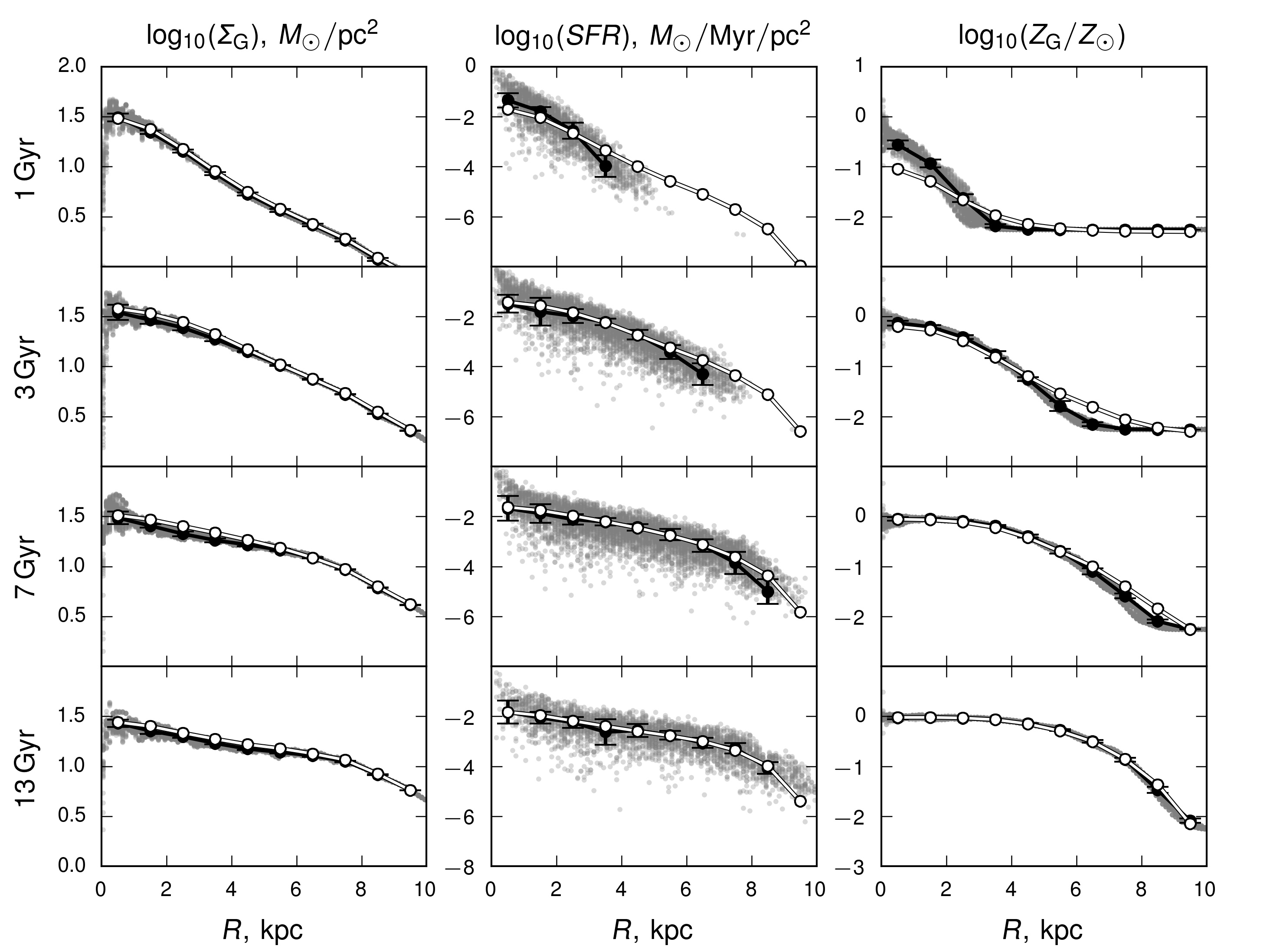,width=120mm,angle=0,clip=}}
\vspace{1mm}
\captionb{6}
{The same as in Fig. 5, but for the HS model.}
}
\end{figure}


\begin{figure}[H]
\vbox{
\centerline{\psfig{figure=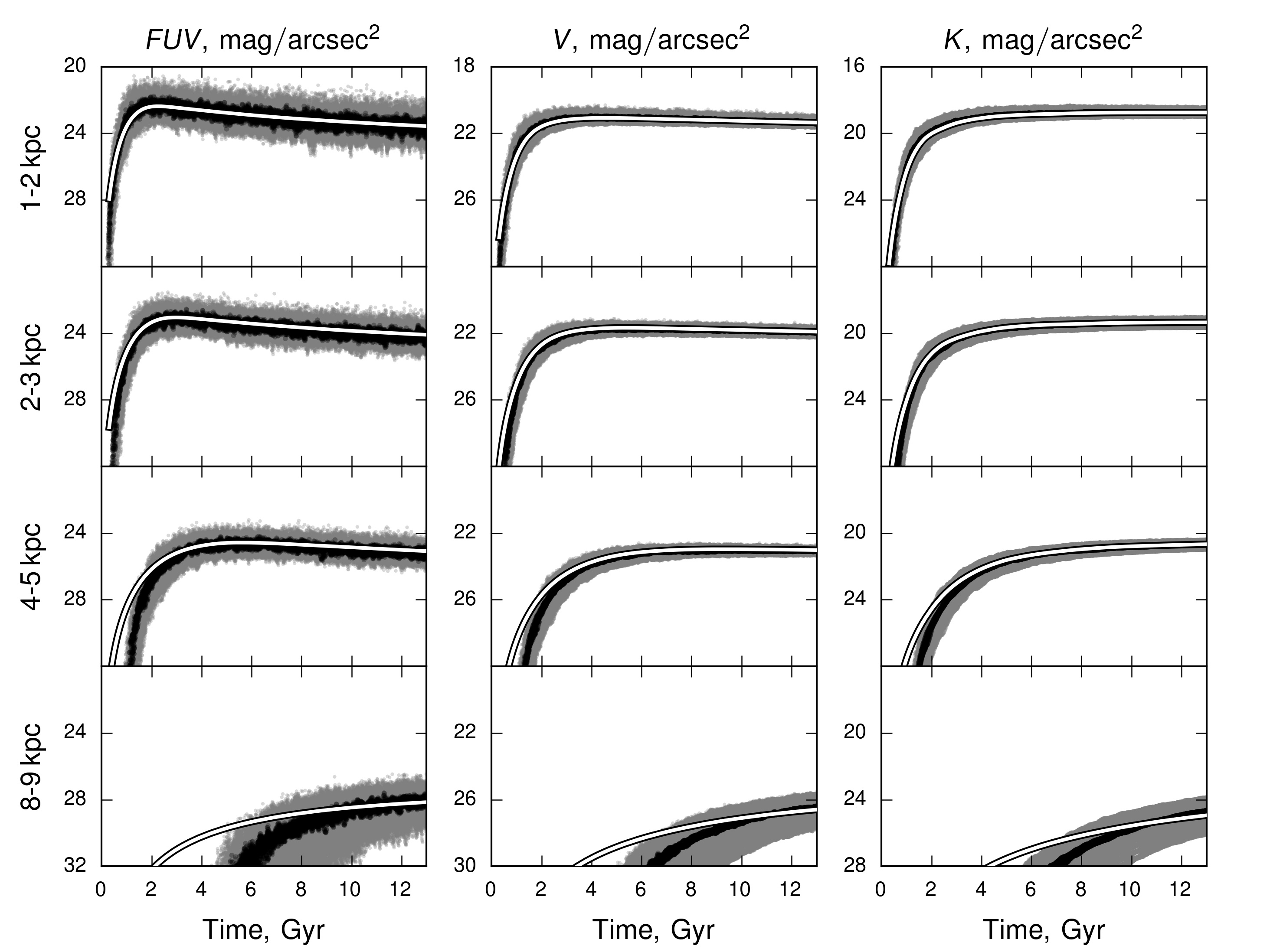,width=120mm,angle=0,clip=}}
\vspace{1mm}
\captionb{7}
{The same as in Fig. 3, but for the surface brightness in the GALEX \textit{FUV}, \textit{V} and \textit{K} passbands.}
}
\vbox{
\centerline{\psfig{figure=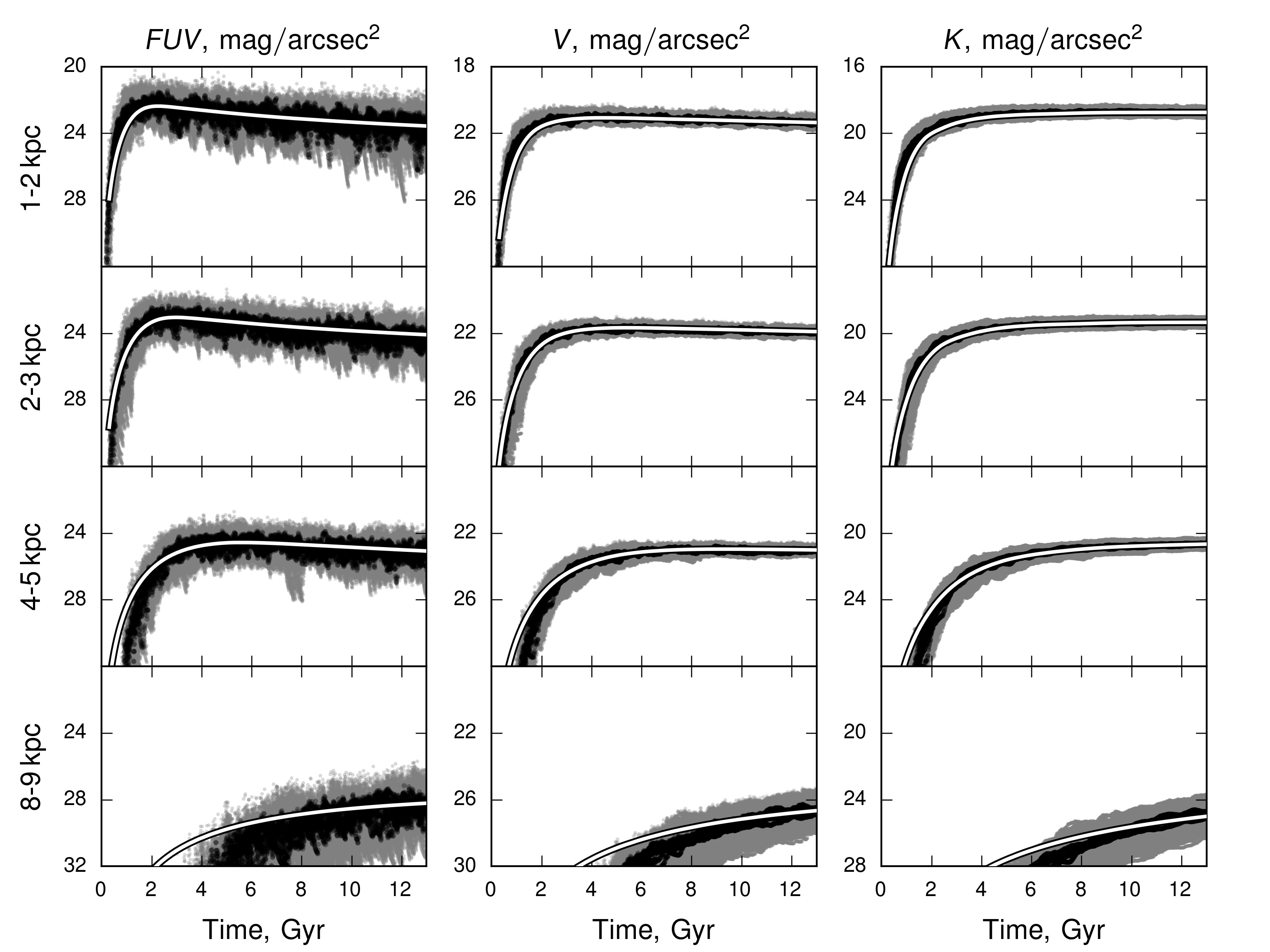,width=120mm,angle=0,clip=}}
\vspace{1mm}
\captionb{8}
{The same as in Fig.\,7, but for the HS model.}
}
\end{figure}

\begin{figure}[H]
\vbox{
\centerline{\psfig{figure=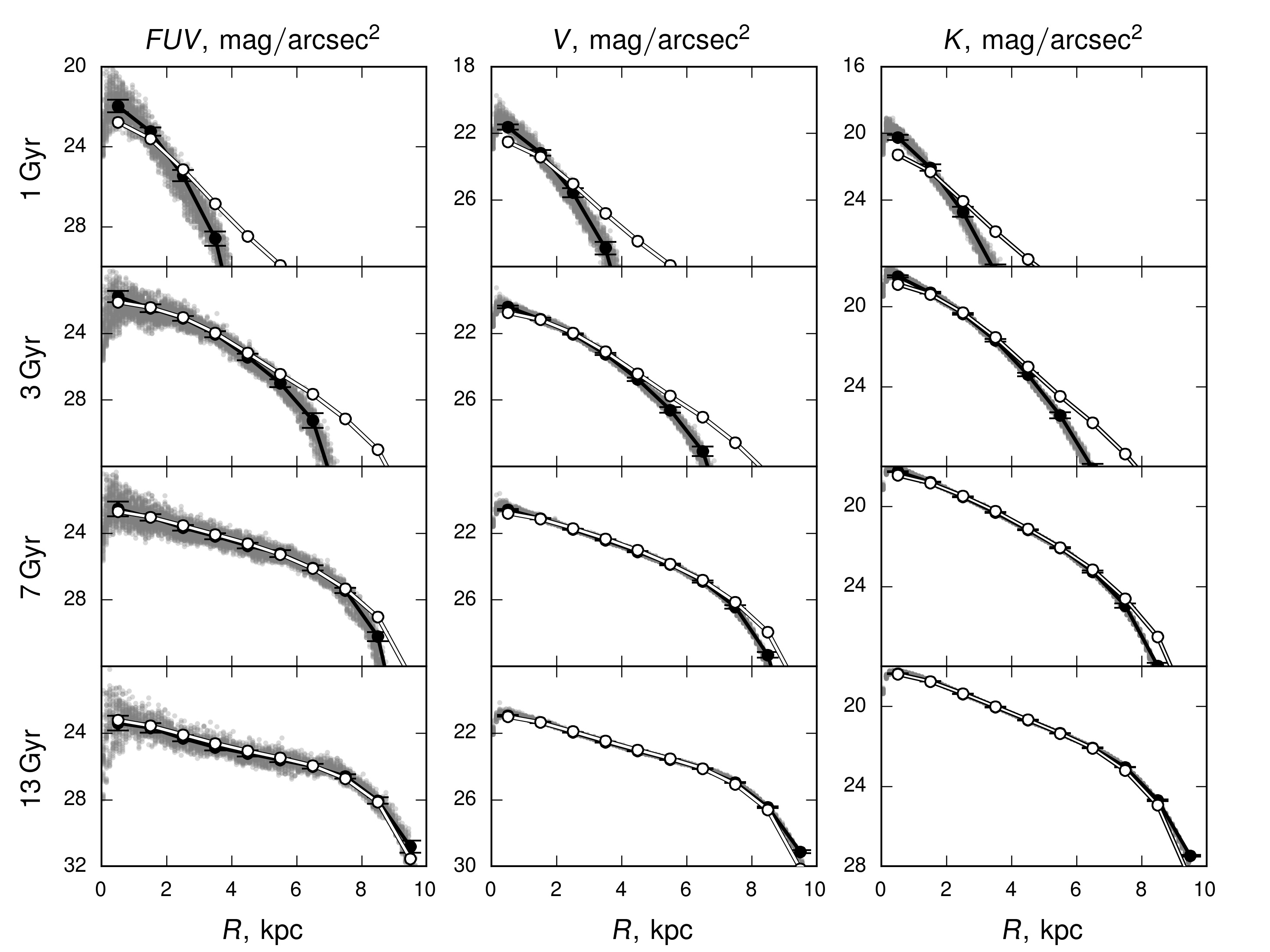,width=120mm,angle=0,clip=}}
\vspace{1mm}
\captionb{9}
{The same as in Fig. 5, but for the surface brightness in the GALEX \textit{FUV}, \textit{V} and \textit{K} passbands.}
}
\vbox{
\centerline{\psfig{figure=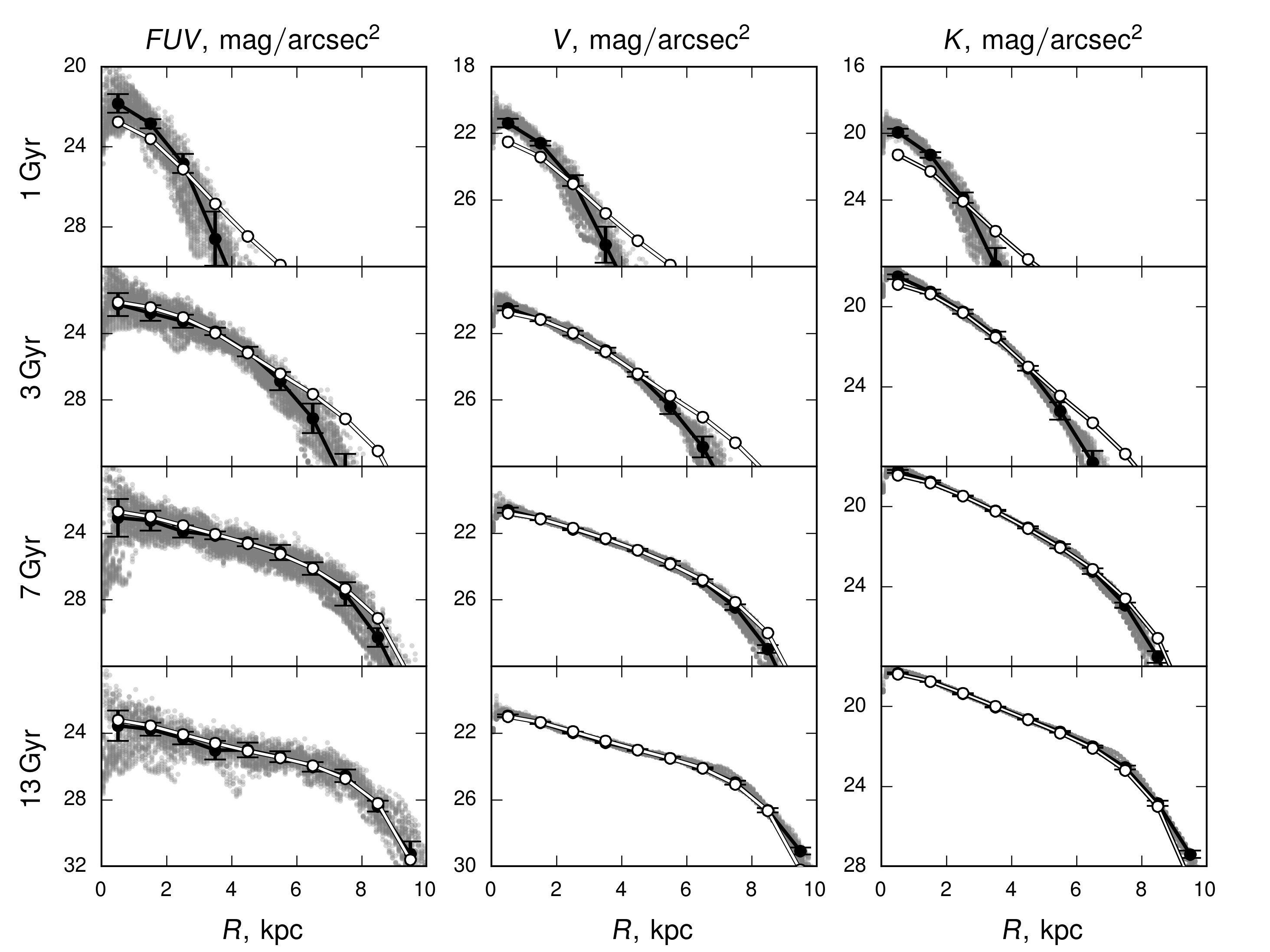,width=120mm,angle=0,clip=}}
\vspace{1mm}
\captionb{10}
{The same as in Fig.\,9, but for the HS model.}
}
\end{figure}


Even stronger stochasticity effects on the SF rate are seen in the 2-D models of high stochasticity (HS), see Fig.~4. The smooth SF 1-D models, shown in all
figures, are calibrated only versus the LS models, because the differences between
the LS and HS calibrations are small.

In order to demonstrate the variations of stochasticity effects along the galaxy radius, we also show the same parameters displayed as radial profiles at different
evolutionary times, see Fig.~5 (LS) and Fig.~6 (HS).

It is well known that spectral energy distributions of stellar populations are
very sensitive to the bursting SF. The evolution of the surface brightness in the GALEX \textit{FUV}, \textit{V} and \textit{K} passbands at different distances from the galaxy center,
averaged within rings 1~kpc wide, are shown in Fig.~7 (LS) and Fig.~8 (HS).
Naturally, the largest stochasticity effects are seen in the GALEX \textit{FUV} passband.
In the outskirts of the disk, however, a large scatter is seen even in the \textit{V} and \textit{K} passbands.

The variations of stochasticity effects along the galaxy radius for the same
passbands, as radial profiles at different evolutionary times, are shown in Fig.~9
(LS) and Fig.~10 (HS). Note, that the stochastic effects almost vanish for the \textit{V}
and \textit{K} passbands at the ages older than 5~Gyr. However, even such small photo-
metric effects of stochasticity can be measured with the present-day observational
technique.

\sectionb{4}{CONCLUSIONS}

We have demonstrated that our stochastic star formation 2-D galaxy models
are, on average, consistent with the smooth star formation 1-D models, described
by the empirical Schmidt-Kennicutt law. However, they show large variations
of star formation rate and surface photometric parameters, especially in the UV
range, across the disk and in time. The largest stochasticity effects are predicted
to occur in the central regions and outer parts of the galaxy disk and during the
first $1$ -- $2$~Gyr from the beginning of the formation of the galaxy.

\thanks{This research was partly funded by a grant No. MIP-102/2011
from the Research Council of Lithuania.}

\References
\refb Barnes K.~L., van Zee L., Dowell J.~D.\ 2013, ApJ, 775, 40
\refb Harris J., Zaritsky D.\ 2009, AJ, 138, 1243
\refb Kang X., Chang R., Yin J. et al.\ 2012, MNRAS, 426, 1455 
\refb Kennicutt R.~C.~Jr.\ 1998, ApJ, 498, 541
\refb Kroupa P.\ 2002, Science, 295, 82 
\refb Le Borgne D., Rocca-Volmerange B., Prugniel P. et al.\ 2004,
A\&A, 425, 881
\refb Magrini L., Corbelli E., Galli D.\ 2007, A\&A, 470, 843 
\refb Marcon-Uchida M.~M., Matteucci F., Costa R.~D.~D.\ 2010,
A\&A, 520, A35 
\refb Mineikis T., Vansevi{\v c}ius V.\ 2010, Baltic Astronomy, 19, 111\
\refb Mineikis T., Vansevi{\v c}ius V.\ 2014, Baltic Astronomy, 23, 209\
\refb Simones J.~E., Weisz D.~R., Skillman E.~D. et al.\ 2014,
ApJ, 788, 12 
\refb Woosley S.~E., Weaver T.~A.\ 1995, ApJS, 101, 181 
\end{document}